\documentclass{article}
\usepackage{graphicx}
\usepackage{scicite}
\usepackage{times}
\usepackage[applemac]{inputenc}
\usepackage{color}
\usepackage{latexsym}
\usepackage{amsmath}
\usepackage{multicol}

\newcounter{lastnote}

\title{Realization of Wheeler’s delayed-choice interference experiment with a single-photon source and space-like separation.}

\author{Vincent Jacques$^{1}$, E Wu$^{1,2}$, Fr\'ed\'eric Grosshans$^{1}$, François Treussart$^{1}$, Philippe Grangier$^{3}$, Alain Aspect$^{3}$, and Jean-François Roch$^{1}^{\ast}$\\
\\
\normalsize{$^{1}$Laboratoire de Photonique Quantique et Mol\'eculaire, ENS de Cachan, UMR CNRS 8537, Cachan, France \\
$^{2}$Key Laboratory of Optical and Magnetic Resonance Spectroscopy, East China Normal University, Shanghai, China\\
$^{3}$Laboratoire Charles Fabry de l’Institut d’Optique, UMR CNRS 8501, Orsay, France}\\
\\
\normalsize{$^\ast$To whom correspondence should be addressed; E-mail:  roch@physique.ens-cachan.fr}
}

\date{}

\begin{document}

\noindent
\begin{center}
{\bf{\sf
\huge
\noindent
Experimental realization of Wheeler's delayed-choice GedankenExperiment

\bigskip
\large
\noindent
V. Jacques$^1$, E Wu$^{1,2}$, F. Grosshans$^1$, F. Treussart$^1$, P. Grangier$^3$, A. Aspect$^3$, and J.-F. Roch$^{3,\ast}$\\
\medskip
\small{%
$^{1}$Laboratoire de Photonique Quantique et Mol\'eculaire, ENS de Cachan, UMR CNRS 8537, Cachan, France \\
$^{2}$Key Laboratory of Optical and Magnetic Resonance Spectroscopy, East China Normal University, Shanghai, China\\
$^{3}$Laboratoire Charles Fabry de l’Institut d’Optique, UMR CNRS 8501, Orsay, France}\\
$^\ast$To whom correspondence should be addressed; E-mail:  roch@physique.ens-cachan.fr
}}

\end{center}

{\sf The quantum “mystery which cannot go away” (in Feynman's words) of wave-particle duality is illustrated in a striking way by Wheeler’s delayed-choice GedankenExperiment. In this experiment, the configuration of a two-path interferometer is chosen after a single-photon pulse has entered it : either the interferometer is \textit{closed} (\textit{i.e.} the two paths are recombined) and the interference is observed, or the interferometer remains \textit{open} and the path followed by the photon is measured. We report an almost ideal realization of that GedankenExperiment, where the light pulses are true single photons, allowing unambiguous which-way measurements, and the interferometer, which has two spatially separated paths, produces high visibility interference. The choice between measuring either the \textit{open} or \textit{closed} configuration is made by a quantum random number generator, and is space-like separated  --- in the relativistic sense --- from the entering of the photon into the interferometer. Measurements in the closed configuration show interference with a visibility of 94\%, while measurements in the open configuration allow us to determine the followed path with an error probability lower than 1\%.}

\begin{multicols}{3}
  Young's double-slit experiment, realized with particles sent one at a time through the interferometer is at the heart of Quantum Mechanics\cite{Feynman}. The striking feature is that the phenomenon of interference, interpreted as a wave following simultaneously two paths, is incompatible with our common sense representation of a particle which implies to follow one route or the other but not both. Several true single-photon interference experiments\cite{Grangier,Jelezko,Zeilinger,Benson,Jacques} have clearly confirmed the wave-particle duality of the lightfield. To understand their meaning, consider the single-photon interference experiment sketched in Fig.1. In the \textit{closed} interferometer configuration, a single-photon pulse is split  by a first beamsplitter $\rm BS_{\rm input}$ of a Mach-Zehnder interferometer and travels through it until a second beamsplitter $\rm BS_{\rm output}$ recombines the two interfering arms. When the phase shift $\Phi$ between the two arms is varied interference appears as a modulation of the detection probabilities at output ports 1 and 2 respectively as $\cos ^{2}\Phi$ and $\sin ^{2}\Phi$. This result is the one expected for a wave, and as Wheeler pointed out, ``\textit{[this] is evidence\ldots that each arriving light quantum has arrived by both routes''}\cite{Wheeler}. If beamsplitter $\rm BS_{\rm output}$ is  removed (\textit{open} configuration), each detector D1 or D2 on the output ports is then associated  to a given path of the interferometer, and --- provided one uses true single-photon light pulses,  ``\textit{[either] one counter goes off, or the other.  Thus the photon has traveled only one route''}\cite{Wheeler}. Such an experiment supports Bohr's statement that the behavior of a quantum system is determined by the type of  measurement performed on it\cite{Bohr}. Moreover, it is clear that for the two complementary measurements considered here, the corresponding experimental settings are mutually exclusive, \textit{i.e.} the beamsplitter $\rm BS_{\rm output}$ cannot be simultaneously inserted and removed. \\
\indent In experiments where the choice between either setting is made long in advance, one could  reconcile Bohr's complementarity with Einstein's local conception of the physical reality. Indeed, when the photon enters the interferometer, it could have received some ``hidden information'' on the chosen experimental configuration and could then adjusts its behavior accordingly \cite{Greenstein}. In order to rule out that too naive interpretation of quantum mechanical complementarity, J. A. Wheeler proposed the ``delayed-choice'' GedankenExperiment in which the choice of which property will be observed is made  after the photon has passed  the first beamsplitter $\rm BS_{\rm input}$. \textit{``Thus one decides the photon shall have come by one route or by both routes after it has already done its travel''}~\cite{Wheeler}.\\
\indent Since Wheeler's proposal, several pioneering delayed-choice experiments have been reported \cite{Alley,Zajonc,Martiennsen,Baudon,Scully,Kawai}. However, none of them fully followed  the original scheme, demanding to use single-particle quantum state and to achieve space-like separation between the choice of the performed measurement and the entering of the particle into the interferometer. We report here the realization of such a ``delayed-choice'' experiment in a scheme very close to the ideal original proposal as described on Fig.1. The choice to insert or remove the beamsplitter $\rm BS_{\rm output}$ is randomly decided, using a Quantum Random Number Generator (QRNG). The QRNG is located close to the output beamsplitter $\rm BS_{\rm output}$, and far enough from the input so that no information about the choice can reach the photon before it passes through the input beamsplitter $\rm BS_{\rm input}$.

Our single-photon source, previously developed for quantum key distribution \cite{Brouri,Alleaume} is based on the pulsed, optically excited photoluminescence of a single N-V color center in a diamond nanocrystal \cite{Kun}. At the single-emitter level, these photoluminescent centers, which can be individually adressed using confocal microscopy \cite{Gruber}, have shown unsurpassed efficiency and photostability at room temperature \cite{Kurtsiefer,Brouri_optlet}. In addition it is possible to obtain single photons with a well defined polarization\cite{Brouri}. (See supplementary information for technical details.)\\
\indent In order to fulfill the requested relativistic separation condition, we use a 48-meter-long polarization interferometer with two spatially separated paths (Fig.2). Single photons, linearly polarized at $45°$ from the eigenpolarizations of the polarizing beamsplitter $\rm BS_{\rm input}$, are split into two components that travel along separated paths. The choice to recombine or not the two paths of the interferometer is done in two steps. First, the two beams, which are spatially separated and orthogonally polarized, are overlapped by a polarizing beamsplitter BS$^\prime$, but can still be unambiguously identified by their polarization. Then, the choice between the two configurations \textit{closed} or \textit{open} is made with the fast electro-optical modulator (EOM) shown on Fig.2, followed by a Wollaston prism. When no voltage is applied to the EOM, the situation corresponds to the removal of BS$_{\rm output}$, and the two paths remain uncombined (\textit{open} configuration of Fig.1). Each detector D1 or D2 is then associated to a specific path, respectively path 1 or path 2. When the EOM half-wave voltage $V_{\pi}$ is applied,  the two orthogonally polarized beams which have traveled along different  paths are recombined by the wollaston prism. We then have the \textit{closed} interferometer configuration of Fig.1.\\
\indent  The space-like separation between the choice of the measurement and the passage of the photon at the input beamsplitter BS$_{\rm input}$ is achieved using a 
homebuilt fast driver capable to switch between  $V=0$ and $V=V_{\pi}$ within 40~ns \cite{Brouri}.  Furthermore, the EOM switching is randomly decided, in real time, by  the QRNG  located close to the output of the interferometer, at 48 meters from BS$_{\rm input}$ (see Fig.2). The random number is generated by sampling the amplified shotnoise of a white light beam. As well known, shotnoise is an intrinsic quantum random process and its value at a given time cannot be predicted \cite{Bachor}. All the experiment, from emission to detection, is synchronized by the clock that triggers the single-photon emission. In particular, in the laboratory frame of reference, the random choice between the \textit{open} and \textit{closed} configurations is simultaneous with the emission of the photon to which that measurement will be applied. In our geometry, it means that the photon enters the future light-cone of that random choice  when it is about at the middle of the interferometer, long after passing BS$_{\rm input}$.

The single-photon behavior is first tested using the two output detectors feeding single and coincidence counters with BS$_{\rm output}$ removed (\textit{open} configuration). We use an approach similar to the one described in References \cite{Grangier} and \cite{Jacques}. We consider a run corresponding to $N_{\mathrm{T}}$ trigger pulses applied to the emitter, with $N_{1}$ (resp. $N_{2}$) counts detected in path 1 of the interferometer by D1 (resp. path 2 by D2), and $N_{\mathrm{C}}$ detected coincidences, corresponding to joint photodetections on D1 and D2 (Fig.2). Any description in which light is treated as a classical wave, like the semi-classical theory with quantized  photodetectors\cite{Lamb}, predicts that these numbers of counts should obey the inequality
\begin{equation}
 \alpha = \frac{N_{C}\times N_{T}}{N_{1}\times N_{2}} \geq 1.
 \end{equation}
 Violation of this inequality thus gives a quantitative criterion which characterizes nonclassical behaviour. For a single-photon wavepacket, Quantum Optics predicts perfect anticorrelation \textit{i.e.} $\alpha = 0$ in agreement with the intuitive image that a single particle cannot be detected simultaneously in the two paths of the interferometer\cite{Grangier}. We measure $\alpha = 0.12 \pm 0.01$ showing that we are indeed close to the pure single-photon regime. The non-ideal value of the $\alpha$ parameter is due to residual background photoluminescence of the diamond sample and to the two-phonon Raman scattering line, which both produce uncorrelated photons with Poissonian statistics.\\
\indent With single-photon pulses in the \textit{open} configuration, we expect each detector D1 and D2 to be unambiguously associated with a given path of the interferometer. To test this point, the ``which-way'' information parameter\cite{Grangierthese,Englert} $I=\lvert (N_{1}-N_{2}) / (N_{1}+N_{2}) \rvert $ is evaluated by blocking one path (e.g. path 2), and measuring the counting rates at D1 and D2. A value of parameter $I$ higher than $0.99$ is measured, limited by detectors darkcounts and residual imperfections of the optical components. Note that the same value is obtained when the other path is blocked (e.g. path 1). In the \textit{open} configuration, we have thus an almost ideal which-way measurement.\\
\indent The delayed-choice experiment itself is performed with the EOM  randomly switched for each photon sent in the interferometer, corresponding to a random choice between the \textit{open} and \textit{closed} configurations. The phase-shift $\Phi$ between the two interferometer arms is varied by tilting the second polarization beamsplitter BS$^\prime$ with a piezoelectric actuator (PZT). For each photon,  we record the chosen configuration, the detection events, and the PZT position. All raw data are saved in real time and they are processed only after a run is completed. For each PZT position, detection events on D1 and D2 corresponding to each configuration are sorted. The results are shown in figure 3. In the \textit{closed} configuration, we observe interference with $0.94$ visibility. The departure from unity  is attributed to an imperfect overlap of the two interfering beams. In the \textit{open} configuration interference totally disappears as evidenced by the absence of modulation in the two output ports when the phase-shift $\Phi$ is varied. We checked that in the delayed-choice configuration, parameters $\alpha$ and $I$ keep the same values as measured in the preliminary tests presented above.

Our realization of Wheeler's delayed-choice GedankenExperiment demonstrates beyond any doubt that the behavior of the photon in the interferometer depends on the choice of the observable which is measured, even when that choice is made at a position and a time such that it is separated from the entrance of the photon in the interferometer by a space-like interval. In Wheeler's words, since no signal traveling at a velocity less than that of light can connect these two events, ``\textit{we have a strange inversion of the normal order of time. We, now, by moving the mirror in or out have an unavoidable effect on what we have a right to say about the already past history of that photon}''\cite{Wheeler}. Once more, we find that Nature behaves  in agreement with  the predictions of Quantum Mechanics even in surprising situations where a tension with Relativity seems to appear \cite{Gisin}. 

\subsection*{Aknowledgements}
 We warmly thank A. Clouqueur and A. Villing for the realization of the electronics of the experiment and J.-P. Madrange for all the mechanical realization of the interferometer. We are grateful to A. Browaeys and L. Jacubowiez for their constant help and many enlighting discussions.
 This work is supported by Institut Universitaire de France.

    \end{multicols}

\pagebreak

        {\sf
\Large
\noindent
\textbf{Figures}}

    \begin{figure}  [h!]
\centerline{\includegraphics[width=9cm]{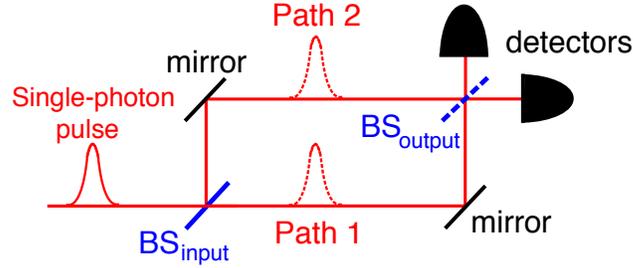}}
  \caption{\it Wheeler's delayed-choice GedankenExperiment with a single-photon pulse in a Mach-Zehnder interferometer. The output beamsplitter BS$_{\mathrm{output}}$ of the interferometer can be introduced or removed (\textit{closed} or \textit{open} configuration) at will. In the \textit{open} configuration, detectors D1 and D2 allow one to unambiguously determine which path has been  followed by the photon. In the \textit{closed} configuration, detection probabilities at D1 and D2  depend on the phase-shift between the two interfering arms. The choice to introduce or remove BS$_{\mathrm{output}}$ is made only after the passage of the photon at the input beamsplitter BS$_{\mathrm{input}}$, so that the photon entering the interferometer ``cannot know'' which of the complementary measurements (path difference vs. which-way) will be performed
  at the ouput.}
\end{figure}

\begin{figure}  [h!]
\centerline{\includegraphics[width=16cm]{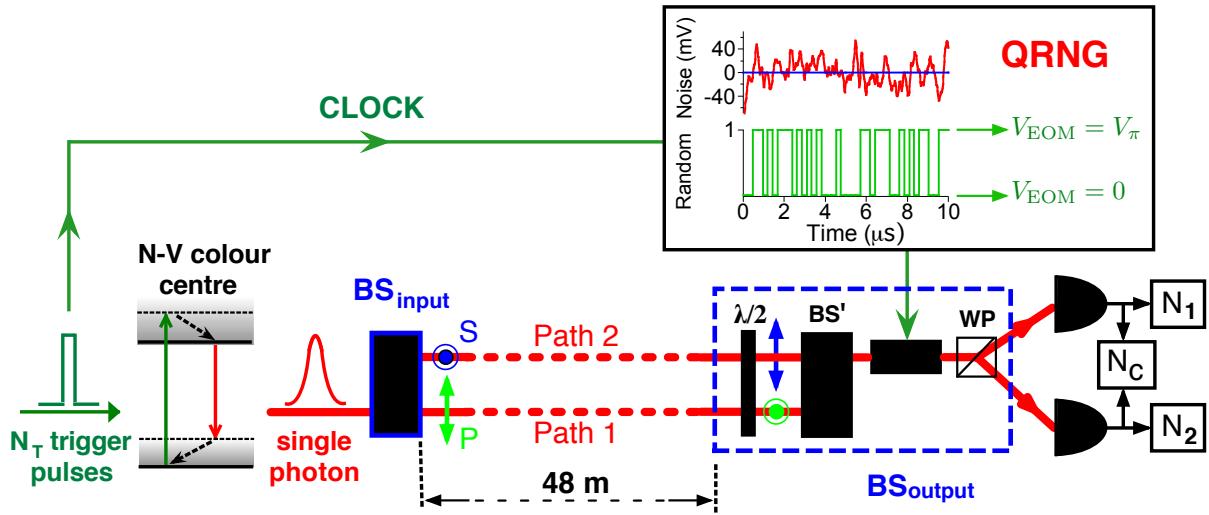}}
  \caption{\it Experimental realization of Wheeler's GedankenExperiment (see Fig.1). Linearly-polarized single-photon pulses, emitted by a single N-V color center, are sent into two spatially separated paths by a polarizing beamsplitter $\rm BS_{\rm input}$. The movable output beamsplitter $\rm BS_{\rm output}$ consists of the combination of a half-wave plate, a polarization beamsplitter BS$^{\prime}$, an electro-optical modulator EOM and a Wollaston prism WP. The choice between the measurement configurations , \textit{open} or \textit{closed}, is realized   by applying a given voltage 0 or $V_{\pi}$ to the EOM. Space-like separation between the entering of the photon into the interferometer and the setting of the chosen experimental configuration, is ensured by locating $\rm BS_{\rm output}$ 48 meters away from $\rm BS_{\rm input}$ and synchronizing the configuration choice with the emission of the single photon (see supplementary information for details). Furthermore, that choice is randomly made accordingly to the output of a Quantum Random Number Generator (QRNG) obtained by sampling the amplified shotnoise from a white light beam, at the clock frequency of the single-photon emission.}
\end{figure}

\begin{figure}  [h!]
\centerline{\includegraphics[width=9cm]{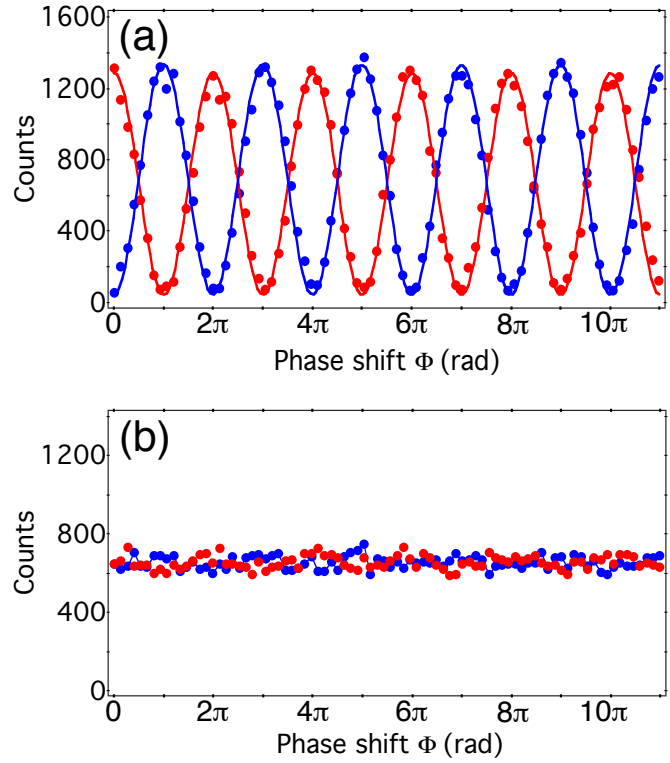}}
  \caption{\it Results of the delayed-choice experiment. The phase-shift $\Phi$ (indicated with arbitrary origin) is varied by tilting BS$^\prime$. Each point, recorded with 1.9~s acquisition time, corresponds to the detection of approximately 2600 photons. Detectors dark counts, $59 \ s^{-1}$ for D1 (blue points) and $70 \ s^{-1}$ for D2 (red points), have been subtracted to the data.  (a) Cases when $V_\pi$ is applied on the EOM (\textit{closed} configuration);  interference with $94\%$ visibility is obtained. (b) Cases when no voltage is applied on the EOM (\textit{open} configuration); no interference is observed and equal detection probabilities ($ 0.50 \pm 0.01$) on the two output ports are measured, corresponding to the washing-out of interference by the which-way information ($I$ parameter larger than $99\%$).}
\end{figure}

\clearpage

    {\sf
\Large
\noindent
\textbf{Methods}}

{\sf
\bigskip
\large
\noindent
\textbf{Triggered single-photon source}
}
\bigskip
\normalsize
\noindent

We use a single nitrogen-vacancy (N-V) color center in a diamond nanocrystal. The N-V centers are created by irradiation of type Ib diamond sample with high-energy electrons followed by annealing at $800$°C\footnote{C. Kurtsiefer, S. Mayer, P. Zarda, and H. Weinfurter, \textit{Phys. Rev. Lett.} \textbf{85}, 290 (2000).}. Under a well controlled irradiation dose, the N-V center density is small enough to allow independent addressing of a single center using standard confocal microscopy\footnote{A. Gruber, A. Dräbenstedt, C. Tietz, L. Fleury, J. Wrachtrup, and C. von Borczyskowski, \textit{Science} \textbf{276}, 2012 (1997)}. The experimental setup used to excite and spectrally characterize single color center photoluminescence is depicted on Fig. 4.

 \begin{figure}  [h!]
 \centerline{\includegraphics[width=14cm]{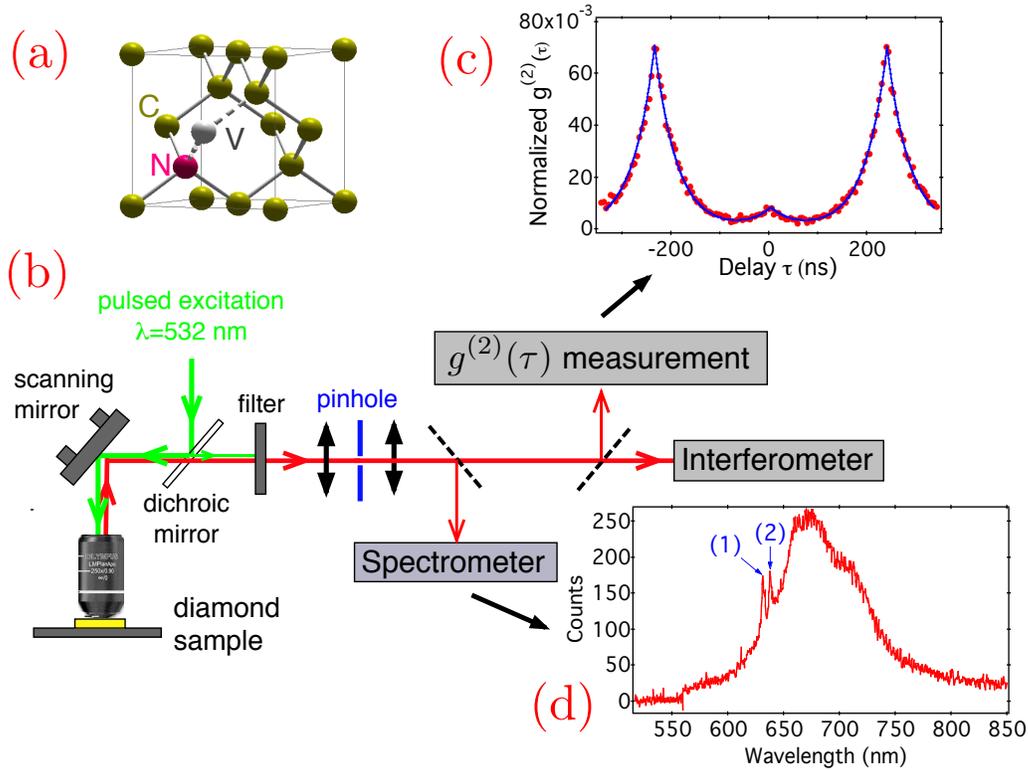}}
  \caption{\it{{\bf (a)-}N-V  color center  consisting in a substitutional Nitrogen atom (N), associated to a
Vacancy (V) in an adjacent lattice site of the diamond crystalline matrix. {\bf (b)-} Confocal microscopy setup. The $532$ nm pulsed excitation laser
beam is tightly focused on a
diamond nanocrystals with a high numerical aperture (NA=$0.95$) microscope objective. The photoluminescence of the N-V color center is collected by the same objective and then spectrally filtered from residual pumping light. Following standard confocal detection scheme, the collected light is focused onto a $100 \ \mu \rm m$ diameter pinhole. To identify a well isolated photoluminescent emitter, the sample is first raster scanned. For the center used in the experiment, a signal over background ratio of about 10 is achieved. {\bf (c)-}The unicity of the emitter is then ensured by observation of antibunching in the second order correlation function $g^{(2)}(\tau)$ of the N-V center photoluminescence, recorded by a standard Hanbury Brown and Twiss setup. The very small remaining value at zero delay $g^{(2)}(0)=0.12$ is due to background emission from the substrate and from the diamond sample in which the color center is embedded. Exponential fit of $g^{(2)}(\tau)$ (blue line) gives the excited level lifetime of the defect  $\tau_{\rm sp}=44.5 \pm 0.5$ ns. {\bf (d)-}Part of the photoluminescence can be also taken to record the emission spectrum of the N-V color center. The two sharp lines (1) and (2) are respectively the two-phonon Raman scattering line of the diamond matrix associated to the excitation wavelength
and the zero phonon line at 637~nm which characterizes photoluminescence of negatively charged N-V color centers.}}
\end{figure}

\indent Excitation is done with a home-built pulsed laser at a wavelength of 532~nm \footnote{A. Beveratos, S. Kuhn, R. Brouri, T. Gacoin, J.-P. Poizat, and P. Grangier,  \textit{Eur. Phys. J. D} \textbf{18}, 191 (2002).}. The laser system delivers 800~ps pulses with energy~50~pJ, high enough to ensure efficient pumping of the color center in its excited level. The repetition rate, synchronized on a stable external clock, is set at $4.2$ MHz so that successive fluorescent decays are well separated in time from each other. Single photons are thus emitted by the N-V color center at predetermined times within the accuracy of its excited state lifetime, which is about 45~ns for the center used in the experiment (see Fig.4-(c)).\\
\indent Significant limitation of defect photoluminescence in diamond arises from the high index of refraction of the bulk material ($n=2.4$), which makes an efficient extraction of the emitted photons difficult. Refraction at the sample interface leads to a small collection efficiency, limited by total internal reflection and strong optical aberrations. An efficient way to circumvent these problems is to consider the emission of defects in diamond nanocrystals, with size much smaller than the wavelength of the radiated light \footnote{A. Beveratos, R. Brouri, T. Gacoin, J.-P. Poizat, and P. Grangier,  \textit{Phys. Rev. A} \textbf{64}, 061802R (2001).}. The sub-wavelength size of the nanocrystals renders refraction irrelevant and one can then simply treat the color center as a point source radiating in air. Furthermore, the small volume of diamond excited by the pumping laser yields very low background light. Such property is of crucial importance for single-photon emission, since residual background light will contribute to a non-vanishing probability of having more than one photon within the emitted light pulse.\\
\indent Nanostructured samples are prepared by starting with type Ib synthetic
diamond powder (ElementSix, The Netherlands) $^{3,4}$. After
irradiation, diamond nanocrystals are dispersed into a polymer solution
and then size-selected by centrifugation, with a mean
diameter of about 90~nm. The resulting polymer solution containing
selected diamond nanocrystals is spin-coated onto the surface
of a dielectric mirror, yielding a 30-nm-thick polymer layer which
subsequently holds the diamond nanocrystals. The ultra-low
fluorescing dielectric structure of the mirror (Layertec, Germany) is
optimized to efficiently reflect the photoluminescence of the
N-V color center towards the collection optics. We note that the background fluorescence from   the mirror
dielectric layers is strongly reduced due to photobleaching after a few hours of sample illumination, while the N-V color center emission properties remain unaffected.

\bigskip
\large
\noindent
\textbf{Single-photon interferometer with two spatially separated paths}

\bigskip
\normalsize
\noindent

The experiment is based on a 48-meter-long interferometer depicted in the article, very close to the Mach-Zehnder interferometer configuration. A linearly polarized single-photon pulse is sent through a first YVO$_{4}$ polarization beamsplitter  (BS$_{\rm input}$) with $45°$ oriented polarization eigenstates. The two S and P linear polarizations at the output of BS$_{\rm input}$ are then spatially separated by 4 mm, sufficient to avoid any overlap between the two beams, since each beam size is about 1 mm. To limit diffraction effects due to open-air propagation along the interferometer, two afocal systems with $\times5$ magnification are used. After $48$~m propagation, equivalent to a time of flight of $\tau_{\rm interf} \simeq 160\rm \ ns$, a second polarization beamsplitter (BS$^\prime$) overlaps the two spatially separated polarizations without recombining the two orthogonally polarized paths of the interferometer. At the output of BS$^\prime$, the two overlapping polarized channels are sent through a KDP electro-optical modulator (EOM, Linos LM0202, Germany) and a Wollaston prism which separates S and P polarizations. Finally, two silicon avalanche photodiodes (APD) operating in the photon counting regime (Perkin Elmer AQR$14$) are positioned at the output ports. Depending on the voltage appplied to the EOM ($V=0$ or $V=V_{\pi}$), the interferometer is either \textit{closed} or \textit{open} as depicted in Fig.5.

\begin{figure}  [h!]
\centerline{\includegraphics[width=11cm]{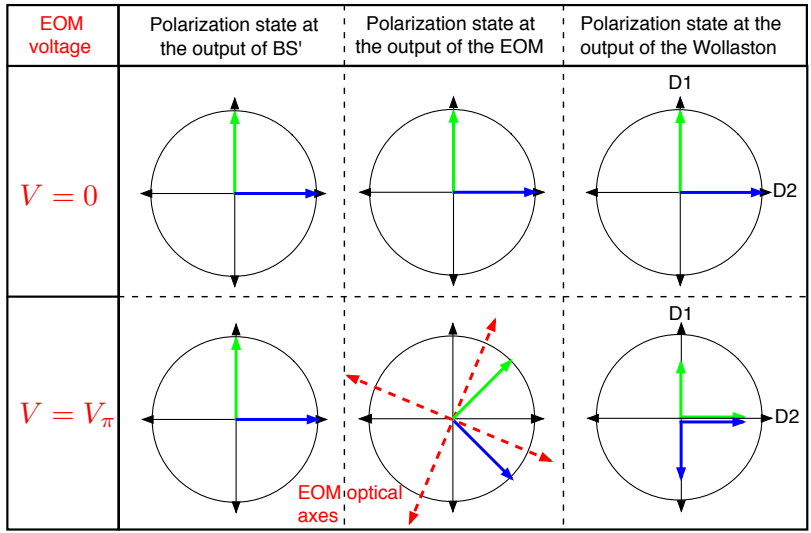}}
  \caption{\it Polarization states after the second polarization beam splitter BS$^\prime$ depending on the voltage applied to the EOM. When no voltage is applied, the two polarizations stay unrecombined and the interferometer is \textit{open}. Detectors D1 and D2, each associated to a given route of the photon along the interferometer, provide a ``which-path'' information. When the $V_{\pi}$ voltage is applied to the EOM, with optical eigenstates oriented at $22.5°$ from the input polarizations, the EOM is equivalent to a half-wave plate which rotates the polarization state by $45°$. The Wollaston prism then mixes the two polarizations and interference appears in the two complementary output ports when the optical path difference between the interfering channels is varied by tilting BS$^\prime$.}
\end{figure}

\indent At last, the N-V center photoluminescence is spectrally filtered with a 10~nm FWHM bandwidth centered at 670~nm to avoid any problem of chromatism of the afocal systems and any reduction of interference visibility due to the broadband emission of the N-V color center (see Fig.4-(d)). Finally counting rates of about $700 \ \rm count.s^{-1}$ are measured on each detector in the \textit{open} configuration. The corresponding signal to noise ratio of about 10 is essentially limited by darkcounts of the two APDs, on the order of $60 \ \rm count.s^{-1}$ for each.

 \bigskip
\large
\noindent
\textbf{Quantum Random Number Generator}

\bigskip
\normalsize
\noindent

 To ensure space-like separation between the entrance of the photon into the interferometer and the choice of the performed measurement, the applied voltage on the EOM is randomly chosen in real time, using a Quantum Random Number Generator (QRNG) located at the output of the interferometer. The random numbers are generated from the amplified shotnoise of a white lightbeam. For each clock pulse, i.e. every $238$ ns, fast comparison of the amplified shotnoise to the zero level generates a binary random number $0$ or $1$. As shown on Fig. 6, the autocorrelation function of a random number sequence reveals no significant correlations between different drafts over the time scale relevant for the experiment. We also checked by direct sampling of the amplified shotnoise every 10 ns that its correlation time is approximatively 60 ns. This measurement confirms that choices made at the $4.2$ MHz clock rate are uncorrelated.
 
\begin{figure}  [h!]
 \centerline{\includegraphics[width=12cm]{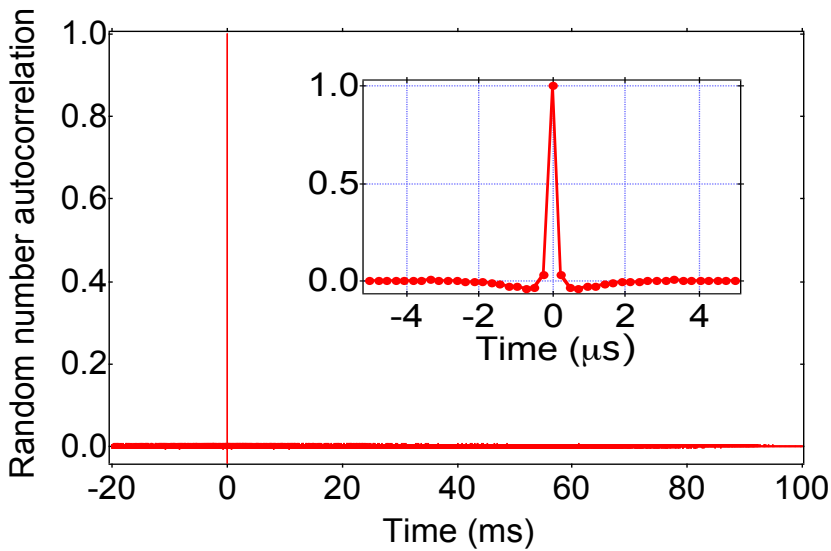}}
\caption{\it{Normalized autocorrelation function of a $420 \, 000$ random numbers sequence generated at the $4.2$ MHz clock frequency. Insert displays a zoom of the function close to zero delay. At long time scale no correlation is observed. A small anticorrelation effect of about $4\%$ appears at very short time scale (below $1\mu \rm s$) presumably due to small oscillations in the amplified output of the shotnoise limited photodetector.}}
\end{figure}

 \bigskip
\large
\noindent
\textbf{Timing of the experiment}

\bigskip
\normalsize
\noindent
A small fraction of the pump pulsed laser at 532~nm  is used to clock-trigger the experiment with 4.2~MHz repetition rate, corresponding to an excitation of the color center every $\tau_{\rm rep}=238 \ \rm ns$ \footnote{Since the time of flight of the photon in the interferometer $\tau_{\rm interf}$ is smaller than the excitation period $\tau_{\rm rep}$, only one single-photon pulse is inside the interferometer at a time.}. As depicted on Fig.7, an FPGA programmable circuit generates for each clock pulse the following sequence. First, fast comparison of the amplified shotnoise to the zero level generates a binary random number 0 or 1 which drives the voltage applied to the EOM, switching between the \textit{open} and \textit{closed} configurations. Then a detection gate of duration $\tau_{\rm d} = 40 \  \rm ns$ is adjusted with appropriate time delays to coincide with the photon arrival on detectors D1 and D2 \footnote{This gated detection leads to a significant decrease of the effective number of dark counts of D1 and D2.}. The FPGA electronics is programmed in order that the random number generation is realized 160~ns before the detection gate, which corresponds to the time of flight $\tau_{\rm interf}$ of the photon inside the interferometer. The QRNG is then drawn simultaneously with the photon emission, within the accuracy of the excited level lifetime $\tau_{\rm sp}$ of the N-V center $\tau_{\rm sp}=44.5 \pm 0.5 \rm ns$ (see Fig.4-(b)).\\

\indent As shown in the space-time diagram of Fig.7, if the single-photon appears at the very beginning (resp. at the very end) of its time-emission window, it has been inside the interferometer for 85~ns (resp. 40~ns), meaning 25~m (resp.12~m) away from the input beamsplitter, when the EOM voltage starts to commute. Furthermore, such timing ensures that the two events ``entering of the photon into the interferometer at BS$_{\rm input}$'' and ``choice of the experimental configuration at BS$_{\rm output}$'' are space-like separated in a special relativistic sense, as required in Wheeler's proposal. Indeed, the photon enters the future light-cone of the random choice  when it is about at the middle of the interferometer, long after passing BS$_{\rm input}$.
\begin{figure}  [h!]
 \centerline{\includegraphics[width=17cm]{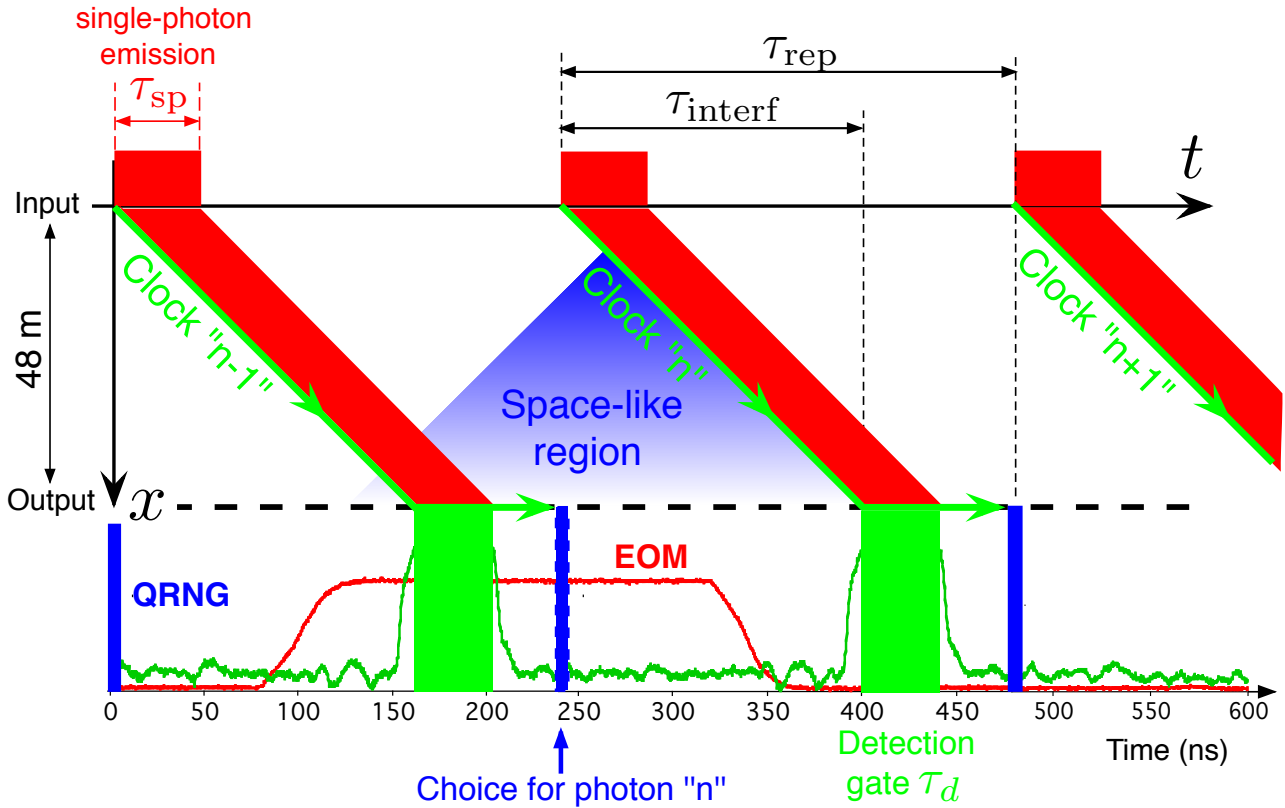}}
\caption{\it{Timing of the ``delayed-choice'' experiment, represented as a space-time diagram. Clock pulses of 5~ns duration are generated by detecting part of the pump laser beam at $532$ nm. Due to the N-V color center radiative lifetime, the single-photon light pulses are emitted during a gate of duration $\tau_{\rm sp}=44.5 \pm 0.5 \rm ns$. All the electronics is based on a FPGA programmable circuit which has a few nanoseconds jitter. To account for propagation delays and response time, the measurement applied to photon ``n'' is synchronised on clock pulse ``n-1'' which triggers the emission of photon ``n-1'' (see green bented line corresponding to speed-of-light propagation). The sequence for the measurement applied to photon ``n'' is done in three steps. First, the binary random number (in blue) which determines the interferometer configuration is generated by the QRNG simultaneously with the trigger of single-photon ``n'' emission. Then, this binary random number (equal to 0 for photon ``n'') drives the EOM voltage  between $V=0$ and $V=V_{\pi}$ within $40$ ns, as shown in the figure in red. Finally the single-photon pulse is detected at the outputs ports by D1 or D2, after its time of flight $\tau_{\rm interf}$ in the interferometer. This detection is done during a gate of duration $\tau_{\rm d}=40$ ns, generated with three electronic D-latches separated by 20~ns (green line). The blue zone represents the space-like domain associated to the event ``entering of photon `n' into the interferometer''. The choice of the open or closed configuration for photon ``n'' is clearly within that region.}}
\end{figure}

\end{document}